      [1999/12/01 v1.4c Il Nuovo Cimento]
\documentclass{cimento}

\usepackage{graphicx}  

\title{UVES - VLT High Resolution Spectroscopy of GRB Afterglows}

\author{S. Piranomonte\from{ins:roma}\ETC\thanks{piranomonte@mporzio.astro.it},
V. D'Elia\from{ins:roma},
P. Ward\from{ins:dub},
F. Fiore\from{ins:roma},
E.J.A. Meurs\from{ins:dub}}

\instlist{
\inst{ins:roma} INAF - Osservatorio Astronomico di Roma, Italy 
    \inst{ins:dub} Dunsink Observatory, Castelknock, Dublin, Ireland
}

\begin{document}

\maketitle

\begin{abstract}

We present early time, high resolution spectroscopy of three GRB
afterglows: GRB050730, 050922C and 060418. 
These data give us precious information on the
kinematics, ionization and metallicity of the interstellar matter of
GRB host galaxies up to a redshift z $\sim 4$, and of intervening
absorbers along the line of sight.

\end{abstract}


\section{Observations}
The GRB050730 afterglow was observed 4.0 hours after the trigger. We find
seven main absorption systems at z=3.968, 3.564, 2.2536, 2.2526,
2.2618, 1.7729 and 1.7723

The GRB050922C afterglow was observed 3.5 hours after the trigger. We find
four main absorption systems at z=2.199, 2.077, 2.008 and 1.9985.

The GRB060418 afterglow was observed 10 minutes after the trigger. We
find four main absorption systems at z=1.489, 1.106, 0.655 and 0.602.

The resolution of all spectra is R 40,000 (7.5 km/s in the observer
frame). Data sets were reduced using UVES pipeline for MIDAS.
All afterglows are clearly detected in the range 3300 - 10000 \AA .

\section{Fine Structure Lines}

Fine structure lines for CII, OI, FeII and SiII have been identified
in all the GRBs. Such lines convey information on the temperature
and electron density of the absorbing medium, provided that they are
excited by collisional processes (J. N. Bahcall, R. A. Wolf et al. \emph{ApJ}, 152, 701, 1968).

\begin{figure} 
  \includegraphics[angle=-90,width=70mm]{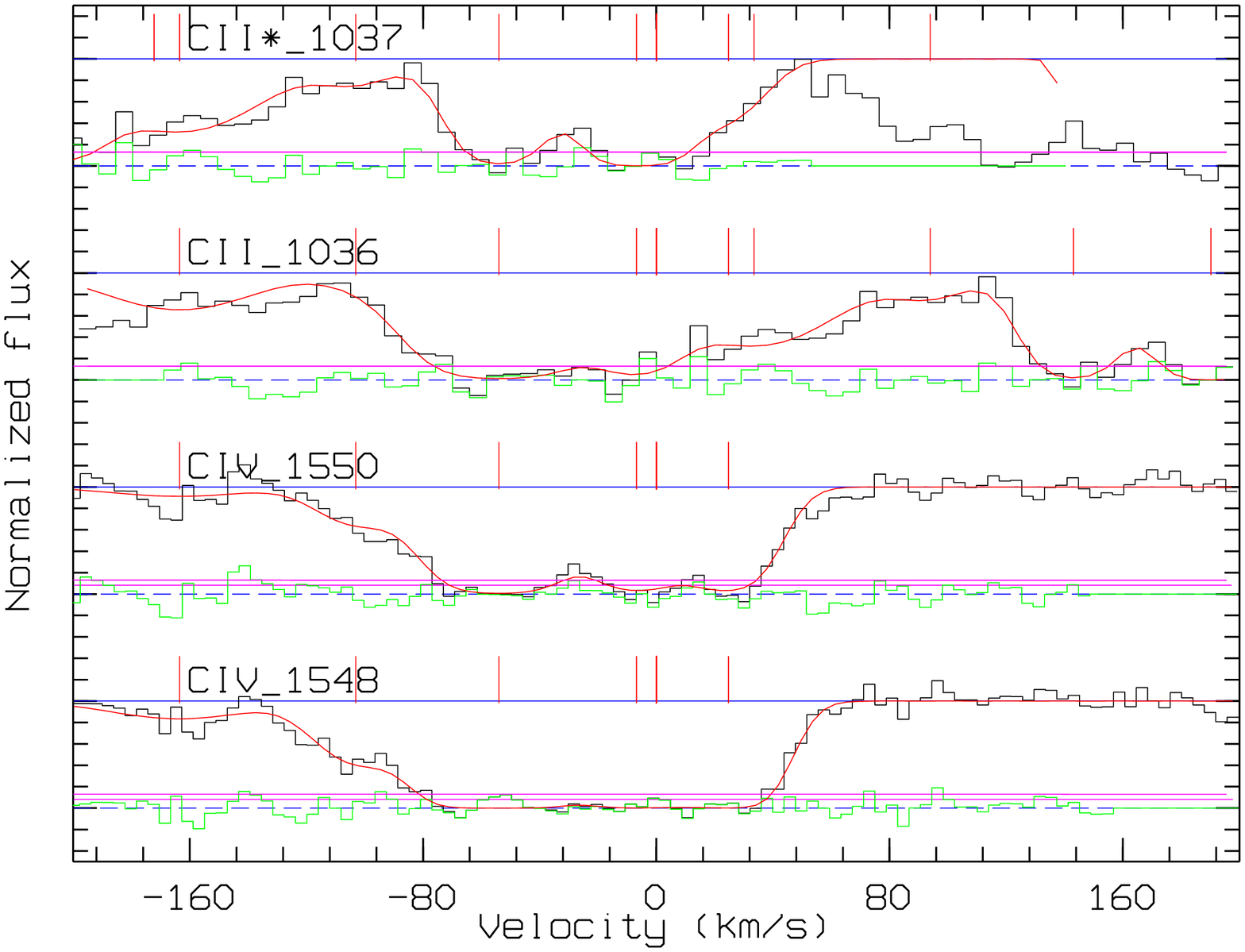}\includegraphics[angle=-90,width=70mm]{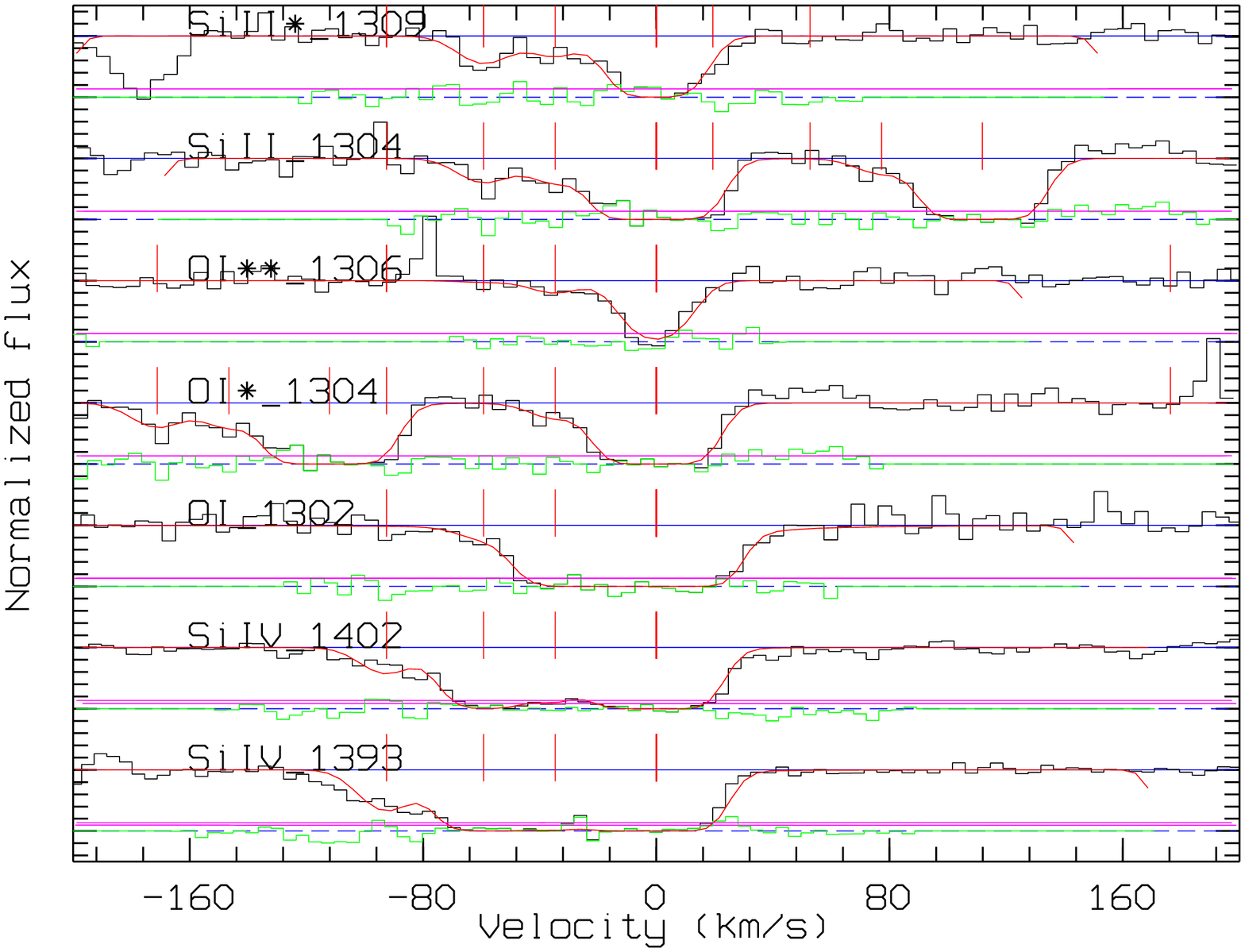}
  \caption{The low ionization lines CII, OI and SiII together with
  their fine structure excited levels in GRB050730. Low ionization
  states and fine structure levels do not appear in all components. High ionization lines CIV and SiIV have been taken as reference lines in the fitting procedure.}
  \label{CIIetc}
\end{figure}

To constrain these parameters we need to estimate the fine structure
column densities for two different ions and compare them. For GRB050730,
two out of five components show fine structure lines
(Fig. \ref{CIIetc}). Reliable values for temperature and electron density are T a
few $10^3$ K and $n > 10^4$ cm$^{-3}$ (second component; the
components are numbered according to decreasing z) and $n \sim 10
\div 100$ cm$^{-3}$ (third component). The other components do not
show fine structure features: this is an indication that they refer to a
clumpy environment.

\section{Metallicity}
Metallicity in GRBs can be measured comparing the column densities of
heavy elements to that obtained for hydrogen by fitting the
Ly$-\alpha$, $\beta$ and $\gamma$ profiles. Both for GRB050730 and
GRB050922C, we find metallicities between $10^{-3}$ and $10^{-2}$ with respect
to the solar values.

Since metals tend to form dust, that then does not contribute to the
absorption lines, this result is affected by some
uncertainties. In GRB060418 we identify CrII and ZnII lines.
Such elements tend to stay in the gas state, minimizing the
uncertainty when estimating the metallicity. No H features are
present in this GRB spectrum, so we derive the $N_{H}$ column from the
X-ray data, leading to: Z(Cr) $= -1.8 \pm 0.3$ and Z(Zn) $= -1.3 \pm0.2$, a
bit higher than for the other two GRBs, but still below the solar
values.

\section{Conclusions}

The absorption spectra of GRB afterglows are extremely complex,
featuring several systems at different redshifts.  Both high and low
ionization lines are observed in the circumburst environment, but
their relative abundances vary from component to component,
indicating a clumpy environment consisting of multiple shells. 

Fine structure lines give information on the temperature and electron
density of the absorbing medium, provided that they are excited by
collisional effects. Different components have different densities,
suggesting a variable density profile.  Metallicity can be derived
from the metal column densities; CrII and ZnII are the best
indicators, since they do not form dust. Metallicity values around $10^{-2}$
with respect to the solar ones have been found.
More details can be found in V. D'Elia, F. Fiore, E.J.A Meurs et al. 2006 (submitted to A\&A, astro-ph/0609825, 2006).


%


\end{document}